# High-Resolution Correlation Spectroscopy of $^{13}$C Spins Near a Nitrogen-Vacancy Center in Diamond

*Abdelghani Laraoui[1], Florian Dolde[2], Christian Burk[2], Friedemann Reinhard[2], Jörg Wrachtrup[2], and Carlos A. Meriles[1,*]*

[1]*Department of Physics, City College of New York – CUNY, New York, NY 10031, USA*

[2]*3rd Physics Institute, University of Stuttgart, 70569 Stuttgart, Germany*

**Spin complexes comprising the nitrogen-vacancy (NV) center and neighboring spins are being considered as a building block for a new generation of spintronic and quantum information processing devices. Because assembling identical spin clusters is difficult, new strategies are in order to determine individual node structures with the highest precision. Here we use a pulse protocol to monitor the time evolution of the $^{13}$C ensemble in the vicinity of a NV center. We observe long-lived time correlations in the nuclear spin dynamics, limited by NV spin-lattice relaxation. We use the host $^{14}$N spin as a quantum register, and demonstrate that hyperfine-shifted resonances can be separated upon proper NV initialization. Intriguingly, we find that the amplitude of the correlation signal exhibits a sharp dependence on the applied magnetic field. We discuss this observation in the context of the quantum-to-classical transition proposed recently to explain the field dependence of the spin cluster dynamics.**

---

[*] To whom correspondence should be addressed. E-mail: cmeriles@sci.ccny.cuny.edu

**Introduction**

Harnessing the quantum behavior of nanoscopic physical systems is at the center of a broad cross-disciplinary effort driven by the promise of various applications for information processing and secure communication[1]. The prospect of performance gains is motivating interest in new devices for existing or new functions — particularly, quantum-based information processing protocols — as well as new paradigms for system architectures beyond CMOS technology. While a dominant technology platform for these applications has yet to emerge, implementations that exploit the quantum properties of individual atomic dopants in a solid-state matrix are of particular importance as this approach has features that make it well positioned to overcome the obstacles to scaling[1,2]. Prominent in this latter category is the electronic spin of the nitrogen-vacancy (NV) center where the low phonon density and the weak spin-orbit coupling typical of carbon structures leads to long spin coherence times under ambient conditions[3]. These isolated spins may be located using confocal microscopy, initialized via optical pumping, and read out through spin dependent photoluminescence measurements[4]. Various alternate architectures for NV-based quantum information processing (QIP) have been proposed, typically relying on local spin clusters comprising one (or more) NV centers as well as ancillary electronic and nuclear spin qubits in its vicinity[5,6]. Flying qubits (e.g., photons) encoded with information on the NV spin alignment are often envisioned as the vehicle to quickly transport information between remote processing units[7,8].

In the absence of schemes to deterministically create identical atomic groupings, the implementation of QIP devices as described above relies on the ability to determine the exact composition and structure of each spin cluster. This information will be important, e.g., to identify weakly coupled nuclear registers in a given spin node[9-11], or to control the set of physical qubits required to back-up the action of a single logical qubit in an error-correction protocol[12].

Here we introduce a new strategy for the characterization of the spin bath in the vicinity of a given NV center. Our approach exploits the comparatively slow evolution of the bath to establish a correlation between the phases picked up by the NV center in contact with the $^{13}$C spins at two different times; upon Fourier transforming the resulting signal, we manage to spectroscopically separate carbon nuclear spins with differing hyperfine coupling constants. Singular about this strategy is that the resolution of the resulting $^{13}$C spectrum is inversely proportional to the spin-lattice relaxation time $T_1$ characterizing the probe spin. Besides the NV center, the latter can be, for example, a long-lived nuclear register, a concept we also demonstrate by making use of the nuclear spin of the host $^{14}$N.

Remarkably, we find that the amplitude of the correlation signal shows a sigmoidal behavior as a function of the applied magnetic field, and abruptly collapses below a critical value $B_c$~10 mT. While the maximum signal amplitude depends strongly on the chosen NV, $B_c$ is found to be approximately the same for all the color centers we investigated thus pointing to well-defined 'high- field' ($B>B_c$) and 'low-field' ($B<B_c$) regimes. These observations suggest a magnetic-field-induced quantum-to-classical transition where the spin dynamics of the quantum system comprising the NV and neighboring nuclear spins suddenly switches at the critical field to a behavior consistent with a simple classical picture. By properly adjusting the time it takes the NV to probe the nuclear Overhauser field, we shift $B_c$ to lower (or higher) values, thus demonstrating our ability to externally control the classicality of this mesoscale spin ensemble at a given external magnetic field.

**Results**

**Probing $^{13}$C spin correlations.** The sample we studied in these experiments is a natural [111] diamond crystal with a nitrogen concentration of approximately 10 ppb. We use confocal microscopy to address individual negatively charged NV centers. These defects have an electron spin triplet ground state ($_3$A) with a splitting $\Delta$=2.87 GHz; optical excitation typically preserves the spin of the $m_S$=0 state whereas the $m_S$=±1 states decay non-radiatively to $m_S$=0 upon undergoing inter-system crossing to a metastable singlet. This mechanism leads to spin polarization of the NV by optical pumping as well as a spin-dependent photon-scattering efficiency that allows for the optical readout of the electronic spin through the photoluminescence intensity[4] (i.e., $m_S$=0 is brighter). In our experiments we apply an external magnetic field $B$ along the [111] axis, which breaks the degeneracy between the $m_S$=±1 states and allows us to address selectively the $m_S$=0→ $m_S$=±1 transitions. For this purpose we use the microwave field created by a thin copper wire overlaid on the crystal surface.

Our approach builds on the notion that when a $^{13}$C coherence is formed, the rate of coherence loss is slow owing to the relatively weak homo-nuclear couplings. Therefore, it should be possible to monitor the bath evolution if we correlate the phases picked up by the NV spin during consecutive interrogations of the bath at times separated by a variable interval. Fig. 1a shows the schematics of our pulse protocol: After optical initialization of the NV center in the $m_S$=0 state, a microwave $\pi$/2-pulse resonant with the $m_S$=0→$m_S$=-1 transition creates a quantum superposition that evolves during a time $2\tau$. We intercalate a $\pi$-pulse at the midpoint of the interrogation period to make the accumulated phase $\phi_1$ insensitive to the spin state of the host $^{14}$N nucleus. We use a 90-degree phase-shifted $\pi$/2-pulse to store a component of the NV magnetization during a time $\tilde{\tau}$, after which we probe the $^{13}$C bath by inducing a second Hahn echo followed by another phase-shifted projection pulse. Denoting with $\phi_2$ the phase accumulated by the NV spin during this second evolution period, one can show that the resulting correlation signal is approximately given by[13]

$$S_{C1}(\tau,\tilde{\tau}) \sim \langle \sin\phi_1 \sin\phi_2 \rangle, \qquad (1)$$

where we use brackets to indicate time average. The subscript is a reminder that in deriving this formula we implicitly assume a (semi) classical view, where $^{13}$C nuclei induce at the NV site a random, time-fluctuating spin-noise field responsible for the accumulated phases. We will later return to this important consideration.

Fig. 1b shows the resulting correlation signal for a given NV center as a function of the time interval $\tilde{\tau}$ separating the first and second interrogation periods of fixed duration $2\tau$; in this particular case, we use an external magnetic field $B$=15.6 mT aligned with the NV axis. We observe a long-lasting response whose amplitude remains virtually unchanged during the first millisecond of evolution. Due to a reduced fluorescence contrast (30% of the maximum possible) we chose to observe only portions of the signal for $\tilde{\tau} > 1$ ms. Figs. 1c and 1d show, nonetheless, well-defined, persisting oscillations with periodicity dominated by the inverse of the $^{13}$C Larmor frequency (6.0 μs) in the applied magnetic field. Also observable are periodic signal beatings, indicative of contributions at different frequencies.

Heuristically, we interpret these observations as the result of a long-lived memory in the spin bath state: When the time interval $\tilde{\tau}+2\tau$ coincides with a multiple of the $^{13}$C Larmor period, the phases picked up by the NV center during both interrogation intervals are the same, thus leading to a (positive) correlation maximum. Conversely, when $^{13}$C spins complete an odd number of half Larmor cycles, $\phi_1$ and $\phi_2$ have different signs, which results in a (negative) correlation minimum. For the particular NV center of Fig. 1 (hereafter labeled as NV1), we find that the correlation envelope decays on a time scale of approximately 4 to 5 ms. We note that this lifetime is comparable to the NV spin-lattice relaxation time $T_1$=6 ms, and an order of magnitude greater than the $T_2$ time as determined from a Hahn-echo protocol (Fig. 1e). Further, the notion that the NV can probe, store, and compare information on the state of the bath at two different times is consistent with the results of Fig. 1f: Here we intercalate a light pulse of variable duration during the bath evolution interval $\tilde{\tau}$. We find that the correlation amplitude progressively vanishes as the green laser pulse re-pumps the NV center into $m_S$=0 prior to the second interrogation period. In this limit and postulating a random phase $\phi_2$, one gets $S_{C1}(\tau,\tilde{\tau}) \sim \langle \sin\phi_2 \rangle \sim 0$ in agreement with our observations.

**Dependence on the external magnetic field.** The experiments above point to a scenario where nuclear spins induce at the NV site a random — though slowly fluctuating — magnetic field ultimately responsible for $\phi_1$ and $\phi_2$. As shown in Fig. 2a, this classical representation of the combined NV-spin bath dynamics breaks abruptly as one brings the external magnetic field below a critical value $B_c$~10 mT. In this regime, we find that the correlation signal vanishes completely regardless the duration of the interrogation periods[14]. Interestingly, the transition takes place over a narrow magnetic field window (~1 mT), thus hinting at mechanisms differing from those invoked in prior studies at variable magnetic field[15,16] where the change is more gradual. Further, the amplitude of the correlation signal varies broadly, from ~40% of the maximum possible contrast to much lower values in the NV set we tested. In particular, some NVs show no correlation signal — at least within our detection limit — for fields up to 40 mT.

The results in Fig. 2a share some similarities with prior studies where coherence is progressively lost as a quantum system controllably couples to a mesoscale open reservoir[17]. In the present case, however, the dynamics are more intricate as it is the spin bath — an open, mesoscale system — the one being interrogated through the quantum response of an individual probe spin. Recent work comparing decoherence in the single- and double-quantum NV transitions upon application of a Hahn-echo protocol provides a more closely related initial platform to interpret the observed behavior[18]. The central idea is that while the combined NV-$^{13}$C ensemble must be generally treated as a closed quantum system, a classical behavior emerges in the limit where the $^{13}$C coupling with the NV is weaker than the nuclear Zeeman energy. In the opposite regime, entanglement permeates the system response thus leading to visibility loss. Within the diamond lattice, the physical boundary between classical and quantum $^{13}$C spins depends on the amplitude of the applied magnetic field. At low fields, the NV decoheres from entanglement with the quantum (i.e., sufficiently close) shell of $^{13}$C spins. As the field increases, this shell progressively shrinks to finally vanish when the applied field exceeds the NV coupling strength to the closest $^{13}$C. The transition from one regime to the other is abrupt since only one strongly coupled carbon is necessary to cause quantum decoherence.

To some extent our results are in agreement with this picture. For example, in Fig. 2d we compare the regime transition for two different echo times $\tau$ chosen to coincide with one half or one fourth the Larmor period of bare $^{13}$C spins $\nu_C$ (black and red dots, respectively). For the shortest interrogation time ($\nu_C\tau = 1/4$), we observe a shift of $B_c$ toward lower values in agreement with the idea that the effect of entanglement on the NV is mitigated by reducing the contact time with the bath. Not shown here for brevity is the converse, namely, $B_c$ shifts to higher values upon increasing $\tau$. This remarkable control over $B_c$ comes at the price of a reduced signal amplitude (see below Fig. 3c); the upper limit is $\nu_C\tau = 1$ where the NV picks up no net phase and the correlation signal disappears (Eq. (1)).

More specifically, for $B > B_c$ and in the limit where nuclear spins are sufficiently far away from the NV, the correlation signal induced by a classical, suitably small field is expected to take the approximate form[13]

$$S_{\text{Cl}}(\tau,\tilde{\tau}) \sim \frac{1}{2}\left(\frac{4\gamma_{\text{NV}} b_{\text{C}}^{\text{rms}}}{2\pi v_{\text{C}}} \sin^2(\pi v_{\text{C}}\tau)\right)^2 \cos(2\pi v_{\text{C}}(2\tau+\tilde{\tau})), \qquad (2)$$

where $\gamma_{\text{NV}}$ denotes the NV gyromagnetic ratio, and $b_{\text{C}}^{\text{rms}}$ is the $^{13}$C root mean square (rms) field. Formula (2) introduces a simultaneous dependence on both the correlation and interrogation times consistent with our observations. An illustration is the case of NV6 (Fig. 3) where alongside the predicted periodic response with frequency $v_{\text{C}}$ (Fig. 3a), we observe the overall signal phase shifts $4\pi v_{\text{C}}\tau$ anticipated in (2) for different Hahn echo times (Fig. 3b). As we change $\tau$, we also observe a concomitant amplitude change with maximum at $v_{\text{C}}\tau = 1/2$, i.e., when $\sin(\pi v_{\text{C}}\tau) = 1$ (Fig. 3c). A vanishing signal is observed as $\tau$ approaches 0 or $1/v_{\text{C}}$ in agreement with the notion that $S_{\text{Cl}}$ must cancel when the NV accumulates a negligible phase (Eq. (1)). Further, the measured dependence is described well by a curve proportional to the fourth power of $\sin(\pi v_{\text{C}}\tau)$ as predicted by (2) (solid line in Fig. 3c). Finally, and given the differing $^{13}$C environments, the rms value of the nuclear field $b_{\text{C}}^{\text{rms}}$ is expected to change broadly from one NV to another, thus explaining the variability of the signal amplitude for fixed $\tau$ (see insert to Fig. 2a).

While the formula above provides the correct periodicity on $\tau$ and $\tilde{\tau}$, numerical integration is generally required to derive the exact curve shape and rms $^{13}$C field at an arbitrary NV site[13]. The problem is that the condition $\gamma_{\text{NV}} b_{\text{C}}^{\text{rms}}/v_{\text{C}} \ll 1$ implicit in (2) is not necessarily valid in most cases, thus rendering the sinusoidal dependence (and overall multiplicative factor) a crude approximation. For example, from the amplitude of the collapses observed in the Hahn-echo signal of NV6 we numerically determine[13] the root mean square (rms) value of the 'classical' $^{13}$C field $\left(b_{\text{Cl}}^{\text{rms}}\right)_{\text{echo}} \approx 4$ µT. Therefore, assuming the 'optimum' contact time $\tau = 1/(2\nu_C)$ in the correlation protocol, we calculate a rms phase $\phi^{\text{rms}} = 2\gamma_{\text{NV}} b_C^{\text{rms}} \tau \approx 2.2\pi$ in a field of 11 m. This value greatly exceeds the 'small angle' range of the NV response (where $|\phi| \ll 1$) and carries a concomitant distortion in the shape of the expected correlation signal. In particular, as $b_C^{\text{rms}}$ increases, the crests and valleys of the sinusoidal response in Eq. (2) gradually give way to positive and negative spikes[13,14]. We plot the calculated correlation signal in Fig. 3a (blue solid line) and find good overall agreement with our experimental observations.

There are some clues, however, pointing to a still incomplete understanding and possibly a more complex, richer spin dynamics. For example, as the applied magnetic field increases, the optimum contact time $\tau = 1/(2\nu_C)$ progressively shortens implying a correspondingly smaller phase $\phi^{\text{rms}}$. A classical $^{13}$C field of amplitude insensitive to $B$ should lead, therefore, to a gradually smaller correlation signal in contrast with our results (Fig. 2a). To shed light on the underlying processes, we use a first-neighbor disjoint cluster model[19] to numerically calculate the quantum evolution of a central NV interacting with a surrounding $^{13}$C bath[13]. In qualitative agreement with Fig. 2, we identify low- and high-field regimes separated by a critical field of about 10 mT. We find, however, that the calculated transition is more gradual and that, in contrast with the high-field plateau of Fig. 2a, the predicted NV response progressively decays after reaching a maximum at about 20 mT (faded red line in Fig. 2a). These findings point to other mechanisms at work not properly modeled in our description of the system dynamics. One possibility that warrants further examination is the gradual dynamic polarization of the $^{13}$C bath, a hypothesis consistent with recent work reporting the observation of large $^{13}$C magnetization in the vicinity of 50 mT[20].

A related question of conceptual importance is whether the observed correlation signal arises from random coherences in the bath, or is rather initiated by the NV itself during the first interrogation period. Only the first scenario — where the bath evolves virtually unperturbed by the NV — is consistent with the semi-classical description above, whereas the second alternative involves some degree of entanglement. Our simulations suggest that both contributions are part of the observed signal though, in principle, one can separate between one or the other provided the $^{13}$C ensemble in the NV vicinity can be polarized and manipulated. Future double-resonance experiments where multiple nuclear spins are controllably initialized into pre-defined pure, superposition, or entangled states will be key to exposing the true classicality of the high-field dynamics.

**High-resolution $^{13}$C spectroscopy.** Not withstanding the exact mechanisms influencing the low- and high-field regimes, a more practical facet of interest is the ability to spectroscopically probe the $^{13}$C ensemble in the NV vicinity. The upper row in Fig. 4 shows the Fourier transform of the correlation signal corresponding to three different NVs. Besides the peak associated to the bath of distant $^{13}$C spins ($v/v_C = 1$), the spectra of NV1 and NV3 bring to light other, closer carbons with various hyperfine couplings. Some of these couplings are strong enough to shift the corresponding resonances beyond the available frequency bandwidth thus leading to folded peaks at deceivingly low frequencies (marked in Fig. 4 with an asterisk).

Similar spectra can be obtained from Fourier-transforming the corresponding Hahn-echo signals of each NV[21], but the relative amplitudes are not necessarily the same. This effect replicates that found in the 3-pulse echo-envelope-modulation protocol[21] and reflects the less-than-optimum impact of hyperfine-shifted carbons at a time $\tau = 1/(2v_C)$ chosen to fit the timing of nuclear spins with near-Larmor resonance frequencies. An interesting example is NV1, whose EPR spectrum reveals a strongly coupled carbon with hyperfine constant of order 600 kHz, substantially higher than the $^{13}$C Zeeman interaction in the applied 11 mT field[13]. From the amplitudes of the corresponding resonances in the echo and correlation spectra, we surmise that this carbon spin plays a relatively minor role on the dynamics of NV1 under the correlation protocol. The latter is consistent with the observation that both NV1 and NV2 share the same critical field $B_c$ even though NV2 does not host any strongly coupled nuclear spin (see Fig. 4c).

In spite of the differing $^{13}$C environments from one color center to another, the time $t_{\text{Corr}}$ during which correlations persist is found to be consistently longer than the typical NV transverse relaxation time, of order 350 μs. The implication is that the ultimate spectral resolution — and, consequently, our ability to discriminate between different $^{13}$C spins in the bath — can be dramatically improved over that possible through the Hahn-echo sequence. For example, in the particular case of NV2, we find $t_{\text{Corr}} \sim 5$ ms (Fig. 1a) comparable to the NV $T_1$, of order 6 ms. Observing carbon evolution over such time intervals is sufficient to resolve hyperfine couplings with a 200 Hz difference, a two-order of magnitude improvement over the resolution attained in prior studies[11].

Fig. 4 introduces a strategy that potentially bypasses the memory loss resulting from NV spin-lattice relaxation. In this case we articulate the protocol of Fig. 1a with a transfer scheme designed to swap the states between the NV center and the host nitrogen nuclear spin immediately after (before) the first (second) interrogation period[13]. During $\tilde{\tau}$ information is stored in the nitrogen nuclear spin (previously polarized to the state $m_I=1$) and $^{13}$C evolution takes place with the NV deterministically initialized to $m_S=1$ or $m_S=0$ (mid and lower traces in Fig. 4d). By comparison to the spectrum without state transfer (upper trace in Fig. 4d) we find an overall reduction of the signal amplitude, which we attribute to the non-ideal fidelity — of order 80%[13] — of our state transfer protocol. More importantly, we find a noticeable change in the relative amplitudes of the hyperfine-shifted resonances resulting from the varying influence of the NV on the bath during $\tilde{\tau}$. For example, when $m_S=0$ the $^{13}$C spectrum reduces to a single peak centered at the bare carbon Larmor frequency, consistent with the notion that the NV-induced magnetic field gradient on the bath is switched off throughout the evolution time.

**Discussion**

Our ability to spectroscopically probe the mesoscale spin system in contact with the NV is a facet of the present technique with important practical implications. Unlike the Hahn-echo or the CPMG protocols, the spectral resolution is defined by the NV $T_1$ time, typically longer than $T_2$. Further, the interrogation time could be extended by storing the phase information in a spin register with a suitably longer spin-lattice lifetime. For example, in the present case the ultimate spectral resolution would scale inversely with the $^{14}$N spin-lattice relaxation time (typically much longer than the NV $T_1$). Naturally, the latter rests on the ability to re-pump the NV to a desired state throughout the full evolution interval so as to prevent NV-induced relaxation of the probed spin bath. This condition is difficult to meet for the 40 mT field chosen for the experiment of Fig. 4 because laser illumination — responsible for the $^{14}$N initialization early in the sequence — has a detrimental effect on the nuclear spin memory when applied during $\tilde{\tau}$ [22]. One can, however, circumvent this problem via modified schemes where initialization of the $^{14}$N spin is carried out by other means at a sufficiently different magnetic field.

The higher spectral resolution and the richer information content of the present technique paves the way to discriminating between otherwise undistinguishable $^{13}$C spins, which would be important to enhance the processing capacity of a given spin node in a QIP device. This is particularly important when we note that several tens of carbon spins (assuming 1% abundance) lie within the 3-4 nm diameter sphere where hyperfine interactions with the NV center are sufficiently strong. The ability to identify and selectively address isolated nuclear qubits within this ensemble will positively impact the number of long-lived registers that can be exploited to store quantum information. Conversely, controlling clusters of interacting carbon spins will be relevant to generating the entanglement required for multiple QIP operations or for implementing quantum error correction protocols. In particular, one can think of improving the fidelity of single-shot readout schemes by concatenating the $^{14}$N spin with other selectively addressable carbons. Alternatively, it would be possible to implement multi-dimensional correlation sequences designed to expose the connectivities between neighboring $^{13}$C spins in ways resembling those used in high-field NMR spectroscopy[23].

We call specific attention to applications in the area of high-resolution sensing to probe spin systems other than the $^{13}$C bath. For example, in a recent paper[24] we have shown that one can use shallow-implanted NVs to probe proton spins within an organic polymeric film deposited on the diamond crystal surface. For this application, the present protocol would lead to a proton 'free-induction-decay' without the need for nuclear spin initialization. The central advantage is versatility: By using radio-frequency pulses resonant with the nuclear Larmor frequency during $\tilde{\tau}$ one could implement homonuclear decoupling sequences to remove dipolar interactions without eliminating the chemical shift information. Alternatively, one can envision double-resonance schemes to implement multidimensional sequences that can shed light on the molecular structure of the organic system at low magnetic field and with nanometer spatial discrimination.

From a more fundamental point of view, the NV-$^{13}$C complex provides an excellent model system to investigate the intricacies of many-body spin dynamics at the mesoscale. Our results show that a relatively small variation of the ratio between the Zeeman and hyperfine energies leads to a dramatic change in the system spin dynamics. Although our results thus far can be qualitatively understood in terms of a field-induced quantum-to-classical transition[18] several features warrant further examination. These include the exact role of strongly coupled nuclei throughout the transition, the system dynamics responsible for the constant signal amplitude at high magnetic field, and the impact of the initial $^{13}$C state on the outcome of the pulse protocol. The critical magnetic field where the regime change takes place can be tuned by adjusting the timing within our correlation scheme, suggesting that a sufficiently short Hahn-echo interrogation can be regarded as a form of weak quantum measurement of the bath state[25]. The latter could possibly be exploited to implement feedback measurement schemes to dynamically protect bath spins from decoherence[26,27].

## Methods

**Optically-detected magnetic resonance.** Our sample is a [111]-oriented, type IIa natural diamond. We use a small permanent magnet in the vicinity of the crystal to induce an external magnetic field $B_0$ aligned with the NV axis. We alter the magnetic field amplitude by displacing the magnet via precision translation stages. We probe NV centers using a purpose-built confocal microscope and a solid-state green laser (532 nm). An ultrafast acousto-optic modulator allows us to create short light pulses with time resolution up to 5 ns. Light is focused by a high-numerical-aperture objective (NA=1.35) on the sample and its fluorescence is collected (after a dichroic mirror and a notch filter) into a 4 μm fiber (whose core serves as the microscope collection pinhole). Photon counting is carried out with the aid of a fiber-coupled avalanche photo detector (APD); the use of a fiber splitter and a second APD in the Hanbury-Brown-Twiss geometry allows us to conduct time-correlated photon counting experiments (necessary to identify the *single* NV centers studied herein). We can obtain scan images of the sample using a two-axis, computer-controlled galvo. We manipulate the NV center spin via resonant microwave pulses (at ~1-4 GHz), which we create through a wave generator and an amplifier, a fast (~3 ns) rf switch and a pulse generator. The field of a 20 μm diameter copper wire positioned over the diamond crystal surface allows us to attain a $\pi/2$ rotation of nearby NV center spins in ~15 ns. Throughout these experiments we selectively address the $m_S=0 \rightarrow m_S=1$ transition in the NV ground triplet. At the magnetic fields we work ($B>1$ mT), all other transitions are strongly detuned and the NV can be effectively considered a two-level system. In the experiments of Fig. 4 we use an arbitrary wave function generator and radio-frequency amplifier to manipulate the nitrogen nuclear spin.

**Acknowledgements**

A.L. and C.A.M. acknowledge support from Research Corporation, the Alexander von Humboldt Foundation, and from the National Science Foundation through grants NSF-1111410 and NSF-0545461. We are thankful to Prof. Glenn Kowach for kindly providing the diamond crystal and are indebted to Jonathan Hodges for his technical assistance with the control software. F.D., C.B., F.R. and J.W. acknowledge financial support by the EU via SQUTEC and Diamant, as well as the DFG via the SFB/TR21, the research groups 1493 "Diamond quantum materials" and 1482 as well as the Volkswagen Foundation. We thank T. Staudacher, J. Michl, and P. Neumann for discussions and support.

**Author contributions**

All authors contributed extensively to the work presented in this paper. F.R. and C.A.M. conceived the initial strategy, and A.L. and F.D. performed the experiments. C.A.M. and C.B. created the analytic model with contributions from F.R. and J.W. J.W. and C.A.M. supervised the project and C.A.M. wrote the manuscript with assistance from all other authors.

**Additional information**

**Competing financial interests**: The authors declare no competing financial interests.

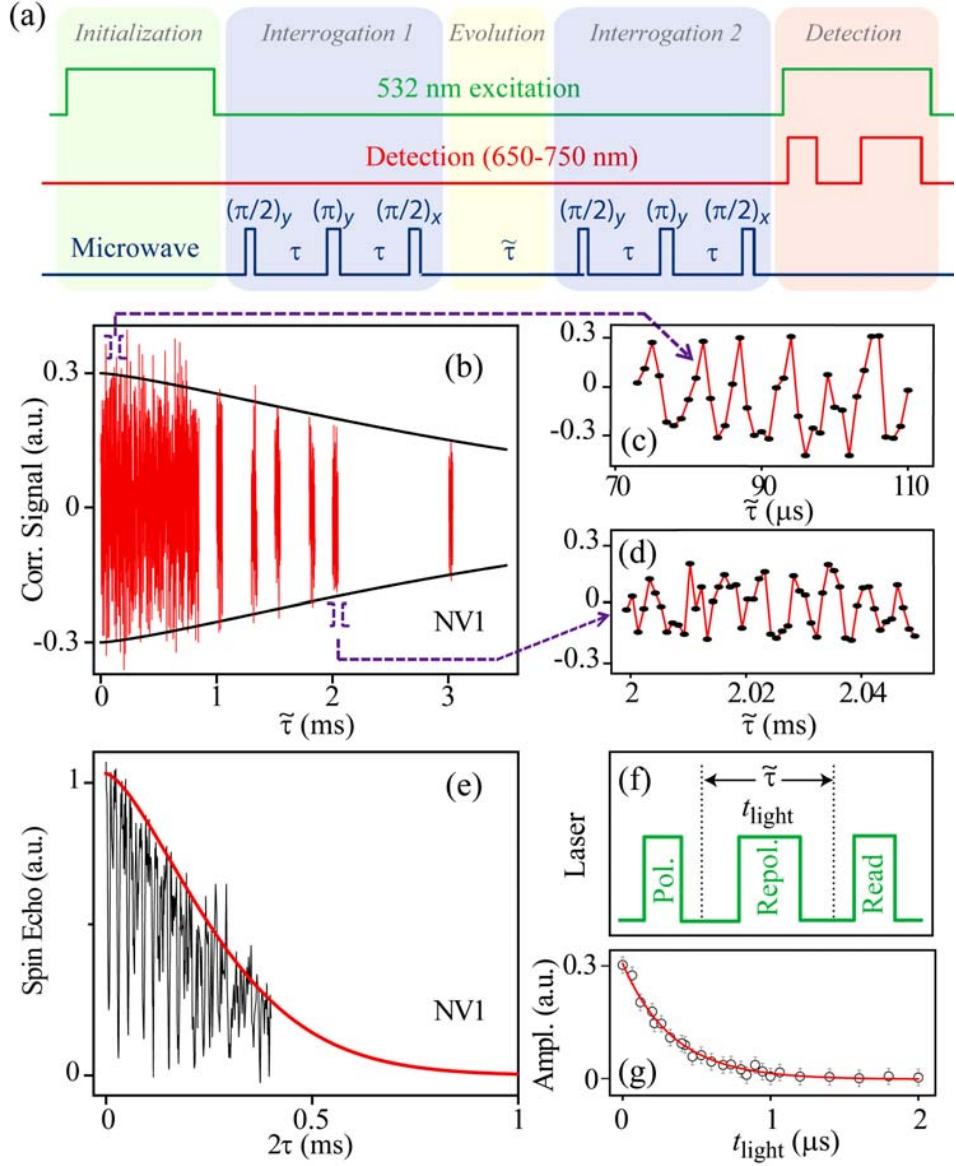

**Fig. 1**: **Monitoring the spin dynamics of the $^{13}$C bath.** (a) Schematics of the pulse sequence. (b) Correlation signal of a defect (labeled NV1 hereafter) in the presence of a magnetic field $B$=15.6 mT; the $^{13}$C Larmor frequency is $v_C$ = 167 kHz. Due to sensitivity limitations, only portions of the signal were measured at times $\tilde{\tau}$ greater than 1 ms. In these experiments $\tau$ is kept constant at 3 µs. (c-d) Expanded views of the signal in (b) at two different time intervals (purple brackets) during the bath evolution period. Correlations persist beyond 3 ms. (e) Hahn-echo signal for NV1 exhibiting a $T_2$ time of ~0.4 ms. (f) Alternate laser protocol with a 'repolarization' light pulse of duration $t_{light}$ intercalated during the evolution period $\tilde{\tau}$. (g) Signal amplitude at short $\tilde{\tau}$ as a function of $t_{light}$. In (b) through (e) all microwave pulses are resonant with the $m_S$=0→ $m_S$=1 transition, and we use a normalized vertical scale (hereafter maintained throughout the text) where -1 or +1 respectively correspond to the fluorescence counts after NV polarization with or without population inversion.

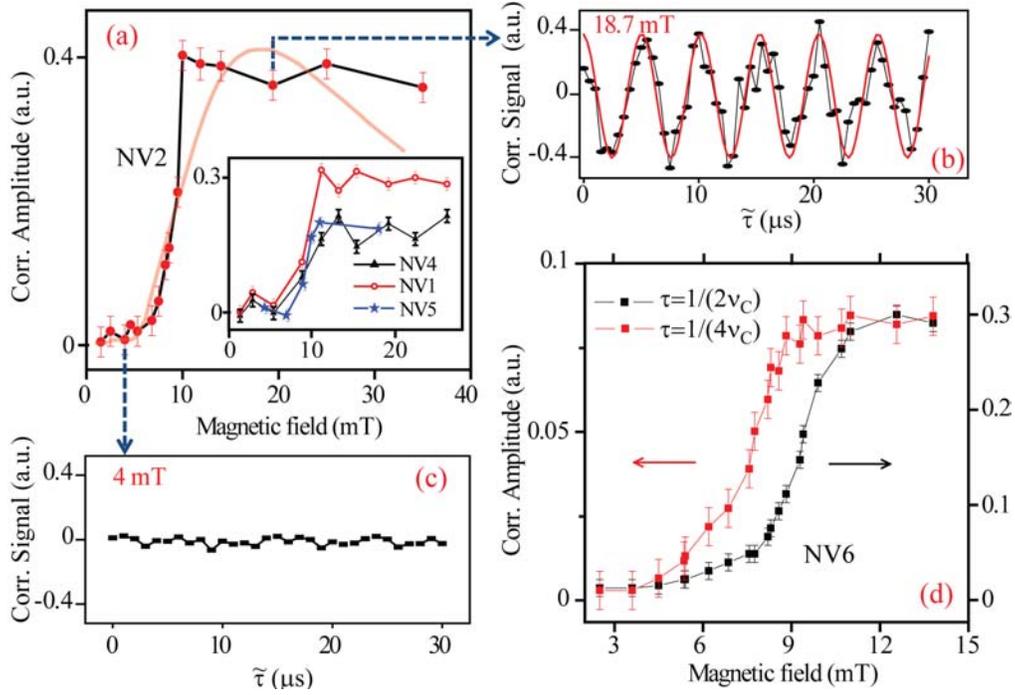

**Fig. 2**: **Magnetic field dependence.** (a) Signal amplitude as a function of the applied magnetic field for NV2 as determined from the system response during the evolution interval $0<\tilde{\tau}<30$ μs. The insert shows similar curves for other NV centers. The fainted solid curve indicates the result from a disjoint cluster simulation (see Ref. [13]). (b) Correlation signal at short $\tilde{\tau}$ for NV2 in an external magnetic field $B$=18.7 mT. The sinusoidal solid line — here serving only as a guide to the eye — exposes the signal periodicity at the $^{13}$C Larmor frequency. (c) Same as in (b) but for $B$=4 mT. In (a) through (c) we choose $\tau=1/(2\nu_C)$. (d) Signal amplitude near the regime change for $\tau=1/(2\nu_C)$ (black dots, right vertical axis) and $\tau=1/(4\nu_C)$ (red dots, left vertical axis). Error bars in (a) and (d) indicate standard deviations.

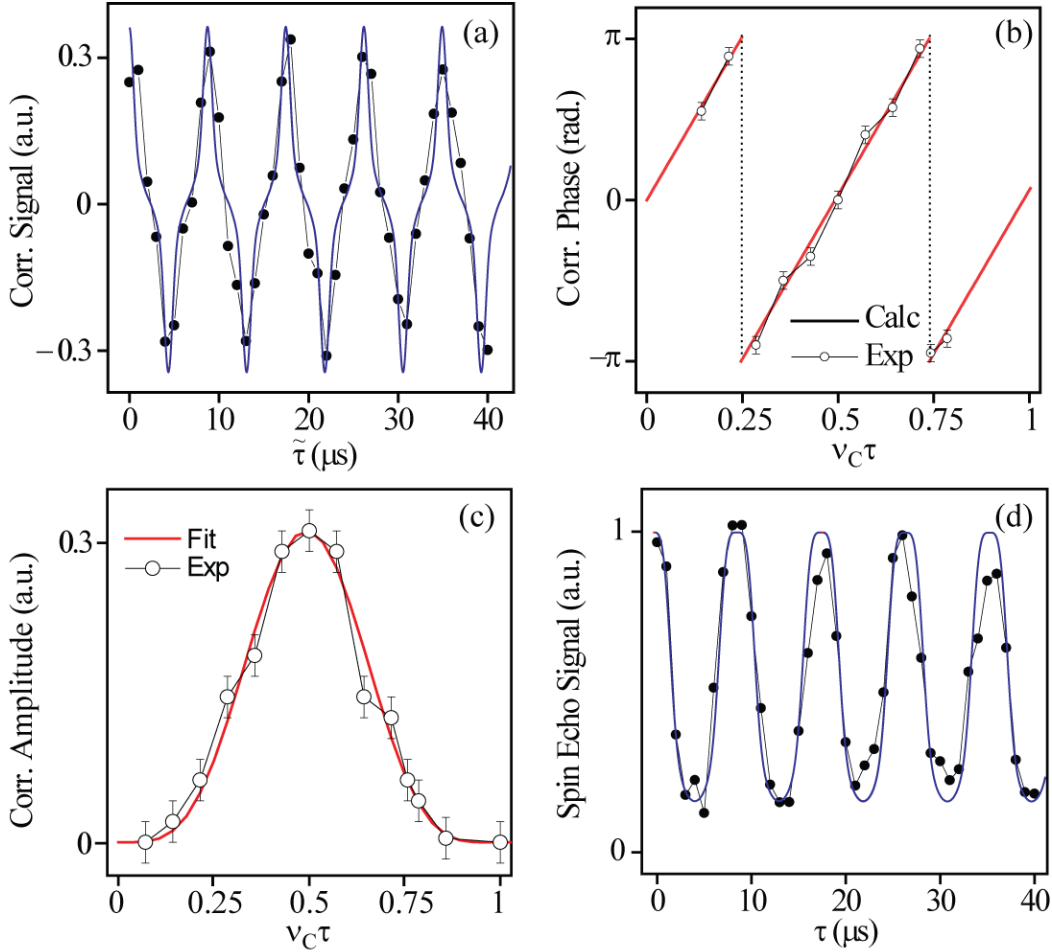

**Fig. 3: Classical response at high magnetic field.** (a) Correlation signal of NV6 at short times $\tilde{\tau}$ in a magnetic field of 11 mT. Solid circles indicate experimental data and the solid curve is the predicted response to a classical random magnetic field of rms amplitude $b_C^{rms}$=4 µT. (b) Measured (open circles) and predicted (solid line) phase shifts in the correlation signal presented in (a) as a function of the normalized echo time. (c) Correlation amplitude at short times $\tilde{\tau}$ (open circles) as a function of $\nu_C \tau$. The solid line is a fit to the function $f(\tau) = A\sin^4(\pi\nu_C\tau)$. (d) Hahn echo signal from NV6 under conditions identical to those in (a). The solid blue line is the calculated response to a classical random field of amplitude $b_C^{rms}$=4 µT. Error bars in (b) and (c) indicate standard deviations.

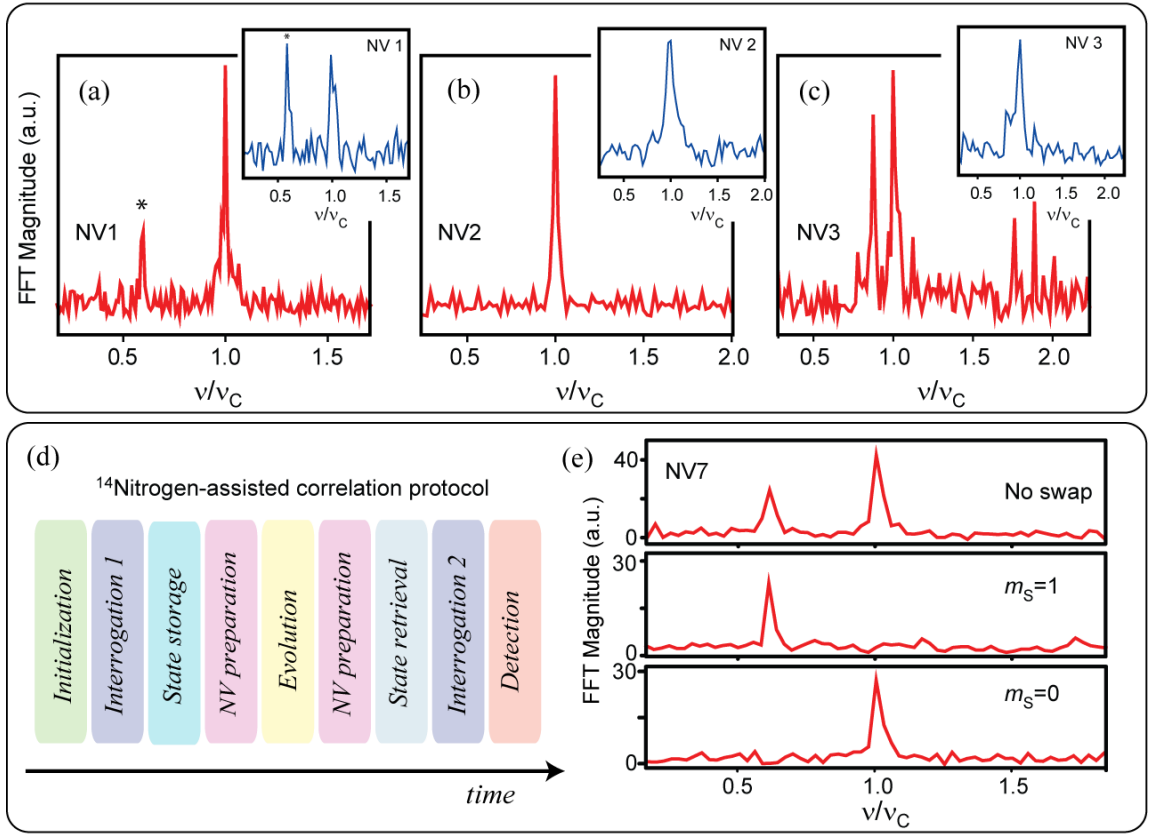

**Fig. 4**: **Spectroscopy of the $^{13}$C spin bath.** (a through c) Fourier transform of the correlation signal for different NVs. The horizontal axis is normalized to the corresponding Larmor frequency $\nu_C$ of bare $^{13}$C spins. In all cases, the Hahn-echo duration $2\tau$ is chosen to coincide with $1/\nu_C$. The inserts show the corresponding Fourier transform of the Hahn echo signal. In (a) an asterisk is used to indicate a peak resulting from spectral folding. The applied magnetic field is $B$=15.6 mT in (a), 16.8 mT in (b), and 18.7 mT in (c). (d) Modified pulse sequence: At 40 mT, laser excitation pumps the NV-$^{14}$N system into the state $m_S$=0, $m_I$=1. A swap gate after the first interrogation period stores the NV state into the $^{14}$N spin and leaves the NV spin in the $m_S$=1 state during $\tilde{\tau}$. An optional, hard $\pi$-pulse brings the NV into $m_S$=0 (NV preparation). After the evolution period, a second swap gate transfers the nitrogen state back onto the NV. (e) Correlation spectrum in the absence of a swap gate (upper trace), and with the NV in state $m_S$=1 (mid trace) or $m_S$=0 (lower trace) during $\tilde{\tau}$. In all three cases we limit $\tilde{\tau}$ to 80 μs.